# DNN-based HRIRs Identification with a Continuously Rotating Speaker Array

Byeong-Yun Ko, Deokki Min, Hyeonuk Nam, and Yong-Hwa Park, *Member, IEEE*

*Abstract*— Conventional static measurement of head-related impulse responses (HRIRs) is time-consuming due to the need for repositioning a speaker array for each azimuth angle. Dynamic approaches using analytical models with a continuously rotating speaker array have been proposed, but their accuracy is significantly reduced at high rotational speeds. To address this limitation, we propose a DNN-based HRIRs identification using sequence-to-sequence learning. The proposed DNN model incorporates fully connected (FC) networks to effectively capture HRIR transitions and includes reset and update gates to identify HRIRs over a whole sequence. The model updates the HRIRs vector coefficients based on the gradient of the instantaneous square error (ISE). Additionally, we introduce a learnable normalization process based on the speaker excitation signals to stabilize the gradient scale of ISE across time. A training scheme, referred to as whole-sequence updating and optimization scheme, is also introduced to prevent overfitting. We evaluated the proposed method through simulations and experiments. Simulation results using the FABIAN database show that the proposed method outperforms previous analytic models, achieving over 7 dB improvement in normalized misalignment (NM) and maintaining log spectral distortion (LSD) below 2 dB at a rotational speed of 45°/s. Experimental results with a custom-built speaker array confirm that the proposed method successfully preserved accurate sound localization cues, consistent with those from static measurement. Source code is available at https://github.com/byko0810/DNN-based-HRIRs-identification.

## I. INTRODUCTION

SPAIAL audio rendering is pivotal for recreating the human perception of spatial audio scenes through headphone or loudspeaker systems. With the emergence of metaverse technologies such as virtual reality (VR) and augmented reality (AR), the demand for spatial audio rendering has surged, aiming to offer users a natural and immersive auditory experience [1]. Moreover, spatial audio rendering finds applications in various fields, including clinical therapy for subjective tinnitus [2], reminiscence therapy [3], auditory localization training for patients with deafness [4], the entertainment audio industry [5], and assistance for the visually impaired [6], highlighting its significance beyond virtual environments.

To simulate spatial audio scenes, the application of head-related impulse responses (HRIRs) to a sound source signal is essential [7]. HRIR represents the time response that characterizes the transmission of sound source from any direction to the ear [8], [9]. Given the substantial variation of HRIRs among individuals, using a non-individualized HRIR for spatial audio synthesis may result in perception errors, such as front-back confusion and inaccurate sound localization [10]. These errors can, in turn, degrade the realism of audiovisual virtual scenes [11].

Various methods have been proposed to obtain individualized HRIRs. One approach involves scanning the head shape and estimating the corresponding individual HRIR through numerical simulation, e.g. finite element method and boundary element method [12], [13]. While these methods have shown promise, particularly at lower frequencies, the complex frequency response of the ear pinna in the higher frequency range poses challenges. Alternatively, deep neural network (DNN)-based HRIR individualization using pinna images or anthropometric parameters has been suggested to derive individual HRIRs [14], [15]. However, training such neural networks effectively requires a large and diverse HRIR dataset that encompasses various head shapes and represents different populations.

Acquiring a HRIR dataset with precise frequency response requires an accurate HRIRs measurement. Given that subjects cannot remain motionless for extended periods, short measurement time is crucial for obtaining HRIRs with high directional resolution [16]. In a conventional static HRIRs measurement, the subject or a speaker array is positioned at a specific azimuth angle, and HRIRs are measured for different elevation angles. The subject or speaker array is then rotated to the next desired azimuth angle, and the process is repeated. This static approach results in considerable measurement time to cover all spherical directions. To speed up this process, the multiple exponential sweep method (MESM) was proposed to mitigate the time required in measuring HRIRs across elevations at a fixed azimuth angle [17]. However, this measurement method still demands considerable time for repositioning the subject or speaker array and completing the sweep signals.

This work was supported by Korea Research Institute of Ships and Ocean engineering a grant from Endowment Project of "Development of Open Platform Technologies for Smart Maritime Safety and Industries" funded by Ministry of Oceans and Fisheries(PES4880). *(Corresponding author: Yong-Hwa Park.)*

Byeong-Yun Ko, Hyeonuk Nam, and Yong-Hwa Park are with Korea Advanced Institute of Science and Technology, Daejeon, South Korea (e-mail: b.y.ko@kaist.ac.kr; frednam@kaist.ac.kr; yhpark@kaist.ac.kr).

Deokki Min is with Korea Institute of Science and Technology, Seoul, South Korea (e-mail: minducky@kist.re.kr).

Color versions of one or more of the figures in this article are available online at http://ieeexplore.ieee.org

Intending to minimize HRIRs measurement time, methods have been suggested measuring HRIRs with a continuously rotating subject or speaker array while excitation signals are emitted from the speaker array [18]. These dynamic approaches substantially reduce HRIRs measurement time and minimizes uncontrollable subject motion during measurement [19]. One of the dynamic approaches is continuous MESM [16], [20], which applies MESM during the continuous rotation of either the subject or the speaker array. However, continuous MESM still requires additional waiting time to account for the overlapped intervals of the excitation signals.

In addition to continuous MESM, analytic model-based dynamic approaches for the identification of time varying HRIRs using white noise signals or perfect sequence/sweep signals simultaneously emitted from the speaker array were suggested [21], [22], [23]. Normalized least mean square (NLMS) algorithm is commonly employed for the analytic model of HRIRs identification due to its robust performance in noisy environments [24], [25], [26]. However, the accuracy of NLMS-based HRIRs identification is degraded as the rotational speed of speaker array or subject increases [27]. Alternatively, Kalman filter has demonstrated enhanced performance in tracking the time varying HRIRs by utilizing the Kalman gain matrix [28], [29]. However, this method necessitates knowledge of noise statistical model, such as the process noise covariance matrix and the measurement noise variance, to accurately conduct HRIRs identification.

To tackle all these problems of HRIRs measurement approaches in terms of both accuracy and speed, this study suggests a novel dynamic approach with DNN-based HRIRs identification with a continuously rotating speaker array. To accurately track the evolution of HRIRs during speaker array rotation, we design the DNN model with fully connected (FC) networks and a Gated Recurrent Unit (GRU) structure combined with sequence-to-sequence learning. The suggested DNN model uses the gradient of the instantaneous squared error (ISE), hidden state, and the speaker excitation signals as inputs to update the HRIRs vector coefficients. To secure the accuracy, we also propose a learnable normalization technique based on the speaker excitation signals to stabilize the scale of the ISE. Additionally, we introduce a whole-sequence updating and optimization scheme that prevents overfitting and enables joint optimization of both HRIRs and the DNN model. We validate the accuracy of the proposed HRIRs identification method through simulations using spatially interpolated HRIRs from the FABIAN HRIR database [30] and experiments conducted on a GRAS 45BC KEMAR Head and Torso Simulator (HATS) with a custom-built rotating speaker array.

The structure of the remaining paper is as follows: Section II defines the problem of time-varying HRIRs from a continuously rotating speaker array. Section III outlines the conventional analytic model-based dynamic approaches for HRIRs identification, followed by Section IV, which introduces our proposed DNN-based HRIRs identification with detailed insights. Sections V and VI present validations through simulations and experiments, accompanied by a thorough analysis and discussion of the results. Finally, Section VII concludes with the efficacy and performance of the proposed method.

## II. TIME VARYING HRIRs FROM A ROTATING SPEAKER ARRAY

The dynamic approaches for HRIRs measurement can be implemented by continuously rotating either a speaker array or a subject. In this study, our aim is to shorten the measurement time by increasing the rotational speed. However, fast rotation of the subject can induce movement and motion sickness [31], [32]. Thus, for both measurement accuracy and subject safety considerations, we focus solely on the scenario involving a rotating speaker array.

In the HRIRs measurement using the continuously rotating speaker array, the sound direction from the speakers undergoes continuous change. The binaural recording captured during the 360° rotation encapsulates the acoustic information of all HRIRs for potential directions. When employing a dense array of closely spaced speakers arranged in a circular configuration along the elevation plane, resulting binaural recording can comprehensively represent the subject's full-spherical HRIRs. By utilizing both the speaker excitation signal and the binaural recording signal, we can derive the complete set of spherical HRIRs. In this setup, the HRIR denotes the impulse response between the speaker of the speaker array and an in-ear microphone placed in the subject's ear. Given that the HRIR continuously evolves during rotation, we need to formulate its behavior over time to analyze the inputs and outputs derived from the dynamic approach for HRIR measurement. Let $x_s(n)$ represent the excitation signal of the $s$-th speaker at time $n$, with the signal length $N$. Additionally, let $h_{n,s}(k)$ denote the $k$-th sample of the time-varying HRIR between the $s$-th speaker and the in-ear microphone in the subject's left or right ear at time $n$. Based on the linear time varying system response [33] and assuming a sampling period in the millisecond range, the output signal measured by the in-ear microphone can be described by:

$$y(n) = \sum_{s=1}^{S}\sum_{k=0}^{K-1} x_s(n-k) h_{n,s}(k) + v(n). \qquad (1)$$

Here, $v(n)$ accounts for measurement noise like white Gaussian noise, modeling the microphone's self-noise under actual conditions. $K$ denotes the length of the HRIR, and $S$ represents the total number of speakers in the array.

To express the time-varying HRIRs of all speakers by vector form, we introduce concise notations. Let $\mathbf{h}_{n,s} = [h_{n,s}(0), \cdots, h_{n,s}(K-1)]$ represent the HRIR vector, encompassing all samples of HRIR for the $s$-th speaker at time $n$. Similarly, $\mathbf{h}_{n,ele} = [\mathbf{h}_{n,0}, \cdots, \mathbf{h}_{n,S}]$ denotes the elevation plane HRIRs vector with $(1, KS)$ shape, consolidating all $\mathbf{h}_{n,s}$ vectors for all speakers. Likewise, $\mathbf{x}_{n,s} = [x_s(n), \cdots, x_s(n-K+1)]$ represents the excitation signal vector for the $s$-th speaker at time $n$, including the $K$ most recent samples. Consolidating all $\mathbf{x}_{n,s}$ vectors across speakers yields an elevation plane excitation signals vector, $\mathbf{x}_{n,ele} = [\mathbf{x}_{n,1}, \cdots, \mathbf{x}_{n,S}]$ and its shape is $(1, KS)$. With these notations, we can concisely rewrite (1) as:

$$y(n) = \mathbf{x}_{n,ele}\mathbf{h}_{n,ele}^{\mathrm{T}} + v(n). \qquad (2)$$

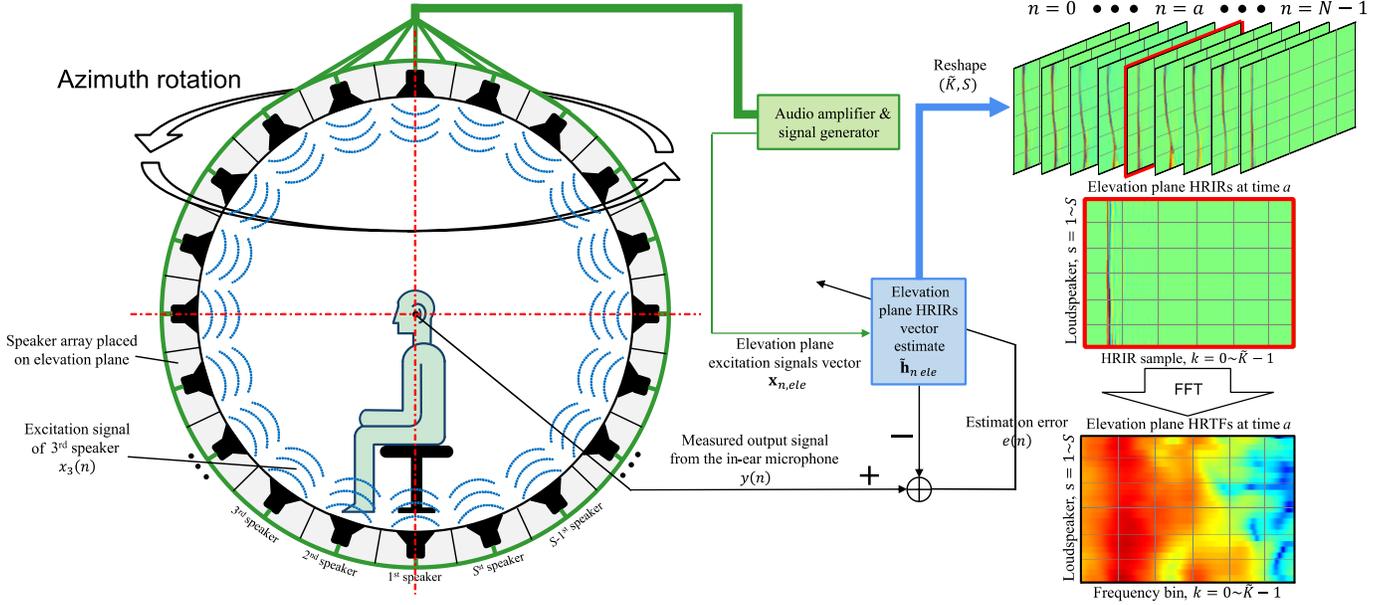

Fig. 1. Overview of the HRIRs identification with an azimuth-rotating speaker array, where the loudspeakers are arranged in a circular configuration along the elevation plane.

The objective of our study is to accurately estimate $\mathbf{h}_{n,ele}$ for each time $n$, and for both the left and right ears. However, in (2), $\mathbf{h}_{n,ele}$ is associated with $\mathbf{x}_{n,ele}$, which contains the excitation signals of all speakers, but is constrained by the single observable output, $y(n)$. As a result, (2) poses an under-determined problem with non-unique solutions and only one constraint, thus making it challenging to derive the time-varying HRIR through conventional deconvolution methods.

### III. HRIRs IDENTIFICATION

Continuous MESM is considered a viable technique for capturing $\mathbf{h}_{n,ele}$, as it involves alternately playing sweep signals from each speaker sequentially with a designated time interval. However, as the number of speakers in the array increases, the measurement time also grows significantly. This is primarily due to the accumulation of time intervals across the numerous speakers, which are repeated for each azimuth angle [16].

To overcome this issue by emitting the signals of all speakers simultaneously, this study adopts system identification, as depicted in Fig. 1. In this process, $\mathbf{h}_{n,ele}$ is predicted by adjusting the coefficient of the elevation plane HRIRs vector estimate, $\tilde{\mathbf{h}}_{n,ele}$, enabling $\tilde{\mathbf{h}}_{n,ele}$ to produce $y(n)$ using $\mathbf{x}_{n,ele}$. This method is referred to as HRIRs identification. Additionally, HRIRs identification employs a perfect sequence [34] or perfect sweep [35] as the speaker excitation signal to prevent correlation between the excitation signals of different speakers and across time $n$.

This approach is built on two key assumptions. First, the time required for HRIR to change during speaker rotation is longer than the duration of the HRIR itself [25]. Second, HRIR exhibits continuity across spatial directions [36]. Regarding the first assumption, HRIR typically has a duration of 23 ms when the distance between the head center and the speaker is less than 1.5 m [37], Moreover, HRIR can be modeled continuously when spatially sampled at intervals of 5° [38]. Therefore, the maximum rotational speed is limited to 217°/s. For the second assumption, HRIR can be expressed using spherical harmonics, which inherently possess spatial continuity properties, as demonstrated in [39].

The shape of $\mathbf{x}_{n,ele}$ and $\tilde{\mathbf{h}}_{n,ele}$ is defined as $(1, \widetilde{K}S)$, where $\widetilde{K}$ represents the estimated length of HRIR. In HRIRs identification, the estimation error $e(n)$ quantifies the difference between the predicted output $\tilde{y}(n)$, derived from $\mathbf{x}_{n,ele}\tilde{\mathbf{h}}_{n,ele}^{\mathrm{T}}$, and the observed output $y(n)$ captured by the in-ear microphone, as follows:

$$e(n) = y(n) - \tilde{y}(n) = y(n) - \mathbf{x}_{n,ele}\tilde{\mathbf{h}}_{n,ele}^{\mathrm{T}}. \qquad (3)$$

To drive $e(n)$ toward zero, the instantaneous square error (ISE) is used as an approximation of the mean square error (MSE). The gradient of the ISE with respect to $\tilde{\mathbf{h}}_{n,ele}$, is given by:

$$\nabla \mathrm{ISE}(n) = \mathbf{x}_{n,ele} e(n). \qquad (4)$$

According to the least mean square (LMS) algorithm, $\nabla \mathrm{ISE}(n)$ is utilized to update $\tilde{\mathbf{h}}_{n,ele}$, resulting in $\tilde{\mathbf{h}}_{n+1,ele}$ as follow:

$$\tilde{\mathbf{h}}_{n+1,ele} = \tilde{\mathbf{h}}_{n,ele} + \mu \nabla \mathrm{ISE}(n), \qquad (5)$$

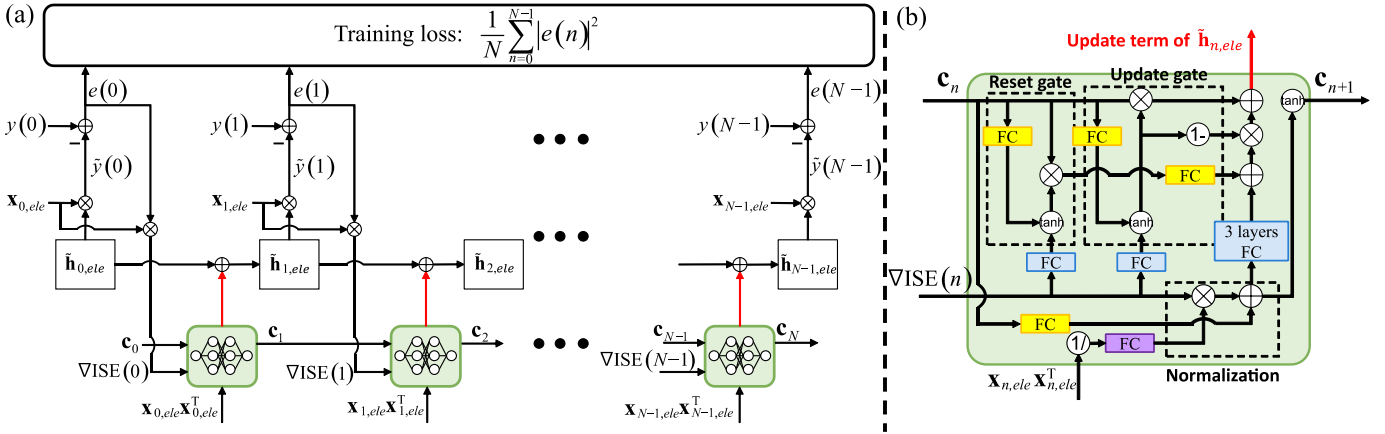

Fig. 2. DNN-based HRIRs identification using our designed DNN model: (a) sequence-to-sequence learning for updating HRIRs vector across time $n$; (b) DNN model and learnable normalization using the power of speaker excitation signals.

Here, $\mu$ is the step-size parameter, which adjusts the update rate of $\nabla\text{ISE}(n)$. By recursively updating $\tilde{\mathbf{h}}_{n,ele}$ for $n = 0$ to $N-1$, the algorithm minimizes ISE, allowing $\tilde{\mathbf{h}}_{n,ele}$ to converge towards $\mathbf{h}_{n,ele}$ after an initial convergence phase [40]. Upon obtaining $\tilde{\mathbf{h}}_{n,ele}$ for all $n$, the final stage of HRIRs identification involves reshaping these vectors from $(1, \widetilde{K}S)$ to $(\widetilde{K}, S)$, as depicted in Fig. 1. These reshaped matrices are then mapped to the azimuth angles corresponding to the speaker array's rotation angles, which are recorded by a gear controller during the measurement process. For instance, $\tilde{\mathbf{h}}_{96000,ele}$ represents the elevation plane HRIRs vector estimate for an 18° azimuth at a rotational speed of 9°/s and a 48 kHz sampling rate, when the rotation of speaker array begins at 0° azimuth.

The LMS algorithm can be enhanced by normalizing the update term, $\nabla\text{ISE}(n)$ using the power of $\mathbf{x}_{n,ele}$, this analytical model referred to as the NLMS algorithm. NLMS is known for its robust performance in system identification within noisy environments [41], owing to its improved stability. However, a limitation of the NLMS algorithm is its lack of a model that accounts for the transition of $\mathbf{h}_{n,ele}$ over time, which may lead to degraded performance when identifying $\mathbf{h}_{n,ele}$ under conditions involving fast rotation of the speaker array [27].

Alternative algorithms, such as recursive least squares (RLS) [42] and Kalman filter [29], can be also applied for system identification. These algorithms utilize analytical models to estimate a system's impulse response and demonstrate rapid convergence. RLS employs a forgetting factor, allowing it to track variations in systems where the impulse response coefficients change over time. However, it has primarily been validated in noise-free environments due to its limited stability [43]. Meanwhile, Kalman filter offers improved tracking of a time-varying system's impulse response through the use of a Kalman gain matrix [28], [29]. Nonetheless, Kalman filter relies on unknown parameters, such as the process noise covariance matrix and the measurement noise covariance matrix, and still faces challenges in accurately capturing $\mathbf{h}_{n,ele}$ variations for the fast speaker rotation [29].

## IV. DNN-BASED HRIRs IDENTIFICATION

In this study, we employ a DNN to leverage its complex architecture for accurately identifying $\mathbf{h}_{n,ele}$. However, a DNN that directly predicts $\mathbf{h}_{n,ele}$ using $\mathbf{x}_{n,ele}$ and $y(n)$ as inputs faces a significant limitation. Specifically, training the DNN requires $\mathbf{h}_{n,ele}$ samples as output labels, which involves measuring multiple subjects with similar physical characteristics to the target subject using the same HRIRs measurement configuration. Creating such a HRIR dataset solely for the identification of one subject's HRIR is highly inefficient and impractical. To address this, we propose DNN-based HRIRs identification, i.e. replacing the analytical model with a DNN to effectively track the dynamic changes in $\mathbf{h}_{n,ele}$ over time.

In related work on DNN-based system identification, Casebeer et al. [44] introduced a method called meta-learning for adaptive filters (Meta-AF). The approach enables the DNN to identify the impulse responses of unknown systems, removing the need for training data of impulse responses. However, it is primarily tailored for stationary systems and still requires a training dataset containing input-output pairs from various system impulse responses for effective training the Meta-AF DNN model.

Given these limitations, a novel DNN model tailored for HRIRs identification is essential. The main objective of this study is to robustly update $\tilde{\mathbf{h}}_{n,ele}$ across $n$ using our designed DNN model, which can be trained without relying on a training dataset, to accurately identify $\mathbf{h}_{n,ele}$. To achieve this, the proposed DNN model utilizes GRU structures to conduct HRIRs identification over a whole sequence as shown in Fig. 2. Additionally, we introduce a normalization, with a learnable form, that uses the power of $\mathbf{x}_{n,ele}$, and establish the training scheme to optimize $\tilde{\mathbf{h}}_{n,ele}$ and the DNN jointly.

The proposed DNN-based HRIRs identification method comprises two main steps:
- Recursive update: the DNN model update $\tilde{\mathbf{h}}_{n,ele}$ recursively for $n = 0$ to $N-1$ using $\nabla\text{ISE}(n)$, the

power of $\mathbf{x}_{n,ele}$, and hidden state. During this process, the ISE is accumulated in the training loss.
- DNN optimization: The accumulated training loss is used to optimize the DNN parameters using a deep learning optimizer.

These two steps are performed sequentially and repeated to ensure accurate identification of $\mathbf{h}_{n,ele}$. For further clarification, a detailed explanation of the DNN-based HRIRs identification is provided below.

*A. Sequence-to-sequence Learning*

In the DNN-based HRIRs identification, the DNN model is utilized to update $\tilde{\mathbf{h}}_{n,ele}$ for $n = 0$ to $N - 1$ as shown in Fig. 2(a), ensuring that $\tilde{y}(n)$ closely approximates $y(n)$, thereby driving $e(n)$ toward zero. Consequently, the DNN performs sequence-to-sequence learning to predict the coefficient vector update for $\tilde{\mathbf{h}}_{n,ele}$ across $n$. To formulate this, we denote the DNN as $g_\phi\{\cdot\}$, parameterized by weights $\phi$. The HRIRs identification process employs an additive update rule as follows:

$$\tilde{\mathbf{h}}_{n+1,ele} = \tilde{\mathbf{h}}_{n,ele} + g_\phi\left\{\nabla\mathrm{ISE}(n), \mathbf{x}_{n,ele}\mathbf{x}_{n,ele}^{\mathrm{T}}, \mathbf{c}_n\right\}, \quad (6)$$

where $\mathbf{c}_n$ represents the hidden state, with shape $(1, S\tilde{K})$, introduced to capture temporal characteristics in the transition of $\mathbf{h}_{n,ele}$. These characteristics arise because $\mathbf{h}_{n,ele}$ is influenced by its previous states $(\mathbf{h}_{n-1,ele}, \mathbf{h}_{n-2,ele}, \mathbf{h}_{n-3,ele}, \cdots)$ due to the continuity across speaker rotation angle. The DNN uses $\nabla\mathrm{ISE}(n)$ as an input to predict the update for $\tilde{\mathbf{h}}_{n,ele}$, leveraging the robustness of gradient descent method for measurement conditions. Our approach focuses on time-domain identification as described in (6), allowing tracking the time varying $\mathbf{h}_{n,ele}$ at each time $n$.

The DNN is aimed to minimize the MSE of $e(n)$ for all $n$. Thus, the optimal weights $\hat{\phi}$ are obtained by solving the following equation:

$$\hat{\phi} = \arg\min_\phi \left\{\frac{1}{N}\sum_{n=0}^{N-1}|e(n)|^2\right\}. \quad (7)$$

When we get $\hat{\phi}$, it ensures that the DNN accurately identifies $\mathbf{h}_{n,ele}$ for all $n$ when applied recurrently to update $\tilde{\mathbf{h}}_{n,ele}$ via (6). During the DNN-based HRIRs identification for the target subject, only the $\mathbf{x}_{n,ele}$ and $y(n)$ sequences, obtained from a binaural recording of the subject using a rotating speaker array, are utilized.

*B. DNN Model*

We design the DNN model based on the recurrent neural network (RNN) as shown in Fig. 2(b) to perform sequence-to-sequence learning and obtain $\tilde{\mathbf{h}}_{n,ele}$ across $n$. However, the length of $n$ is far long. For example, if the sampling rate is 48 kHz and rotational speed is 45°/s, 192000 sequences are composed of $n$ for 180° rotation. Such a long sequence is challenging for standard RNNs, which often suffer from the vanishing gradient problem, limiting their ability to accurately identify $\mathbf{h}_{n,ele}$ over the whole sequence. To address this, we incorporate GRU structures, which alleviate the vanishing gradient issue while minimizing the number of DNN parameters, thereby reducing computational cost. In our model, the GRU's reset and update gates are applied along with a one-layer fully connected (FC) network and a tanh activation function. This configuration helps regulate the balance of information between past and present time steps, allowing gradients to flow through the whole sequence and enabling effective tracking of long sequences of $\mathbf{h}_{n,ele}$. In addition, we employ 3-layers FC network to convert $\nabla\mathrm{ISE}(n)$ to the coefficient vector update for $\tilde{\mathbf{h}}_{n,ele}$ at the output layer of the DNN model. This is based on the findings in previous work [29]. The work showed that using a single linear matrix (e.g., Kalman gain) to convert $e(n)$ into an update term of $\tilde{\mathbf{h}}_{n,ele}$ fails the identification at fast rotational speed due to the variations of acoustical features in $\mathbf{h}_{n,ele}$ caused by the head and torso's reflection and diffraction. Therefore, we utilize 3 dense layers on the FC network to enable the accurate modeling of the complex $\mathbf{h}_{n,ele}$ transitions.

During the sequence-to-sequence learning, the scale of $\nabla\mathrm{ISE}(n)$ varies significantly, particularly in the initial convergence phase when $e(n)$ is large. The DNN does not utilize $\mu$, making it unable to regulate the update rate, potentially causing divergence of $e(n)$ and preventing the sequence-to-sequence learning. The normalization method in NLMS addresses this by dividing $\nabla\mathrm{ISE}(n)$ by the power of $\mathbf{x}_{n,ele}$, projecting $\nabla\mathrm{ISE}(n)$ onto $\mathbf{x}_{n,ele}$ [22]. However, the fixed normalization term and update rate can lead to slow convergence or degraded performance [45]. We extend this method by introducing a learnable normalization to adaptively normalize $\nabla\mathrm{ISE}(n)$ and control the update rate, achieving fast convergence and accurate tracking of $\mathbf{h}_{n,ele}$. This is done by including $\mathbf{x}_{n,ele}\mathbf{x}_{n,ele}^{\mathrm{T}}$ as an additional input to the DNN model and adding two one-layer FC networks at the input stage. The first FC network processes the reciprocal of $\mathbf{x}_{n,ele}\mathbf{x}_{n,ele}^{\mathrm{T}}$ to produce a normalization vector of shape $(1, S\tilde{K})$. This vector is then applied element-wise to $\nabla\mathrm{ISE}(n)$. The second FC network processes $\mathbf{c}_n$ and combines its output with the normalized $\nabla\mathrm{ISE}(n)$ as shown in Fig. 2(b). Together, these networks adaptively adjust the scale of $\nabla\mathrm{ISE}(n)$ and the update rate depending on $n$. The output of this process is used in two ways: it is passed to the three-layer FC network to compute the coefficient vector update for $\tilde{\mathbf{h}}_{n,ele}$, and simultaneously fed through a tanh activation function to generate the next hidden state $\mathbf{c}_{n+1}$.

*C. Whole-sequence Updating and Optimization Scheme*

An effective training method is essential to ensure that the designed DNN updates $\tilde{\mathbf{h}}_{n,ele}$ to closely approximate $\mathbf{h}_{n,ele}$ across $n$. However, due to the absence of a training dataset, the DNN can only be optimized using information derived from the $\tilde{\mathbf{h}}_{n,ele}$ updating process. To address this, we propose a tailored training scheme for the DNN-based HRIRs

**Algorithm 1:** Whole-sequence updating and optimization.

**function** TRAIN
   $epoch \leftarrow 1$          ▷Initialize epoch
   $\phi \leftarrow \text{init}$          ▷Initialize DNN parameter
   $\mathbf{x}_{ele} \leftarrow [\mathbf{x}_{0,ele}, \mathbf{x}_{1,ele}, \ldots, \mathbf{x}_{N-1,ele}]$    ▷Concatenate inputs
   $\mathbf{y} \leftarrow [y(0), y(1) \ldots, y(N-1)]$    ▷Concatenate outputs
   $\tilde{\mathbf{h}}_{0,ele}, \mathbf{c}_0 \leftarrow \mathbf{0}, \mathbf{0}$    ▷Initialize updating process
   **while** $epoch$ **not** 300 **or** $L_{opt}$ **not** CONVERGED **do**
      $\tilde{\mathbf{h}}_{ele}, L \leftarrow \text{UPDATE}(g_\phi, \mathbf{c}_0, \tilde{\mathbf{h}}_{0,ele}, \mathbf{x}_{ele}, \mathbf{y})$
      $L_{train} \leftarrow \ln(L/N)$       ▷Training loss (8)
      $\nabla L_{train} \leftarrow \text{GRAD}(L_{train}, \phi)$    ▷Gradient of training loss
      $\phi \leftarrow \text{ADAM}(\phi, \nabla L_{train})$    ▷DNN optimization
      $epoch \leftarrow epoch + 1$       ▷Update epoch
   **return** $\phi$
**function** UPDATE$(g_\phi, \mathbf{c}_0, \tilde{\mathbf{h}}_{0,ele}, \mathbf{x}_{ele}, \mathbf{y})$
   $L \leftarrow 0$          ▷Initialize loss
   **for** $n \leftarrow 0$ to $N-1$ **do**
      $\tilde{y}(n) \leftarrow \mathbf{x}_{n,ele} \tilde{\mathbf{h}}_{n,ele}^T$    ▷Predicted output
      $e(n) \leftarrow y(n) - \tilde{y}(n)$    ▷Estimation error
      $\text{ISE}(n) \leftarrow \|e(n)\|^2$    ▷ISE
      $\nabla \text{ISE}(n) \leftarrow \mathbf{x}_{n,ele} e(n)$    ▷Gradient of ISE
      $\xi(n) \leftarrow \{\nabla \text{ISE}(n), \mathbf{x}_{n,ele} \mathbf{x}_{n,ele}^T, \mathbf{c}_n\}$    ▷DNN inputs
      $(\Delta_n, \mathbf{c}_{n+1}) \leftarrow g_\phi\{\xi(n)\}$    ▷DNN outputs
      $\tilde{\mathbf{h}}_{n+1,ele} \leftarrow \tilde{\mathbf{h}}_{n,ele} + \Delta_n$    ▷Update HRIR
      $L \leftarrow L + \text{ISE}(n)$    ▷Accumulate loss
   $\tilde{\mathbf{h}}_{ele} \leftarrow [\tilde{\mathbf{h}}_{0,ele}, \tilde{\mathbf{h}}_{1,ele} \ldots, \tilde{\mathbf{h}}_{N-1,ele}]$    ▷Concatenate HRIRs
   **return** $\tilde{\mathbf{h}}_{ele}, L$

identification.

A major difficulty arises from the fact that an unoptimized DNN produces inaccurate identification $\tilde{\mathbf{h}}_{n,ele}$, which in turn can lead to exploding loss and hinder effective DNN optimization. To tackle this issue, the proposed training scheme alternates between updating $\tilde{\mathbf{h}}_{n,ele}$ and optimizing the DNN, allowing joint optimization of $\tilde{\mathbf{h}}_{n,ele}$ and the DNN. During the update phase, the DNN update $\tilde{\mathbf{h}}_{n,ele}$ across $n$, while the ISEs are collected for each time step. In the subsequent optimization phase, these collected ISEs are used to refine the DNN's parameters, resulting in improved updates of $\tilde{\mathbf{h}}_{n,ele}$. For this optimization step, standard deep learning optimizers can be employed, such as stochastic gradient descent (SGD) [46] or adaptive moment estimation (Adam) [47]; in this study, the Adam optimizer is used. Through this iterative training scheme, both the DNN and $\tilde{\mathbf{h}}_{n,ele}$ are progressively improved. To ensure the DNN generalizes across all the whole sequence and avoids overfitting to a limited time range, optimization is performed only after the DNN has completed updating $\tilde{\mathbf{h}}_{n,ele}$ for the whole sequence, $n = 0$ to $N-1$. This scheme—referred to as the whole-sequence updating and optimization scheme—ensures consistent identification across all time $n$ steps.

The detailed procedure of the whole-sequence updating and optimization scheme is as follows:

1. $\tilde{\mathbf{h}}_{0,ele}$ and $\mathbf{c}_0$ are initialized as zero vectors with shape $(1, S\tilde{K})$. At the beginning of the updating process, $e(0)$, ISE(0), and $\nabla$ISE(0) are calculated using $\mathbf{x}_{0,ele}$, $y(0)$, and $\tilde{\mathbf{h}}_{0,ele}$. $\nabla$ISE(0), along with $\mathbf{x}_{0,ele} \mathbf{x}_{0,ele}^T$, and $\mathbf{c}_0$ are fed into the DNN, which updates $\tilde{\mathbf{h}}_{1,ele}$ and $\mathbf{c}_1$.

2. The DNN recursively updates $\tilde{\mathbf{h}}_{n,ele}$ and $\mathbf{c}_n$ across $n$ until $N-1$. During this process, the DNN parameters remain fixed, and ISE($n$) is accumulated across $n$ to form the training loss.

3. After updating $\tilde{\mathbf{h}}_{n,ele}$ for the whole sequence, the accumulated loss is used to compute the training loss $L_{train}$. The DNN is optimized by computing the gradient of training loss using standard deep learning tools, including backpropagation through time (BPTT), with the Adam optimizer.

The combination of accumulated loss across all $n$ and BPTT allows the DNN to effectively learn HRIRs identification for the whole sequence. Steps 1, 2, and 3 are repeated iteratively until the optimizer loss converges or until a maximum of 300 epochs is reached, where one epoch consists of executing all three steps once. For initialization of the DNN, the fully connected (FC) layers within the DNN are set as identity matrices. A simplified version of the training algorithm is presented in Algorithm 1, where GRAD denotes the gradient of the first argument with respect to the second.

To facilitate the whole-sequence updating and optimization scheme, the training loss is designed to promote accurate updating of $\tilde{\mathbf{h}}_{n,ele}$ across the whole sequence. Specifically, we utilize the average ISE, computed as the mean squared error (MSE) of $e(n)$ for $n = 0$ to $N-1$, as the training loss. Additionally, based on previous research on DNN-based system identification [44] applying the natural logarithm to the MSE of $e(n)$ reduces the dynamic range of the training loss, thereby enhancing the robustness of DNN training for system identification. The training loss is thus defined as:

$$L_{train} = \ln\left(\frac{1}{N}\sum_{n=0}^{N-1}\|e(n)\|^2\right). \tag{8}$$

In addition, a small learning rate of $10^{-4}$ and a large momentum term of 0.9 are employed to prevent issues such as the exploding gradient problem commonly encountered in sequence-to-sequence learning.

To validate the proposed DNN-based HRIRs identification, we evaluate the algorithm through simulations and experiments, as detailed in Sections V and VI, respectively. We compare its performance against the conventional analytic models and Meta-AF. Section V includes an ablation study of the proposed method, while Section VI applies a fixed configuration for experimental validation.

## V. SIMULATION

### A. Simulation Set-up

We conducted simulations to validate the proposed DNN-based HRIRs identification. The FABIAN HRIR database [30]

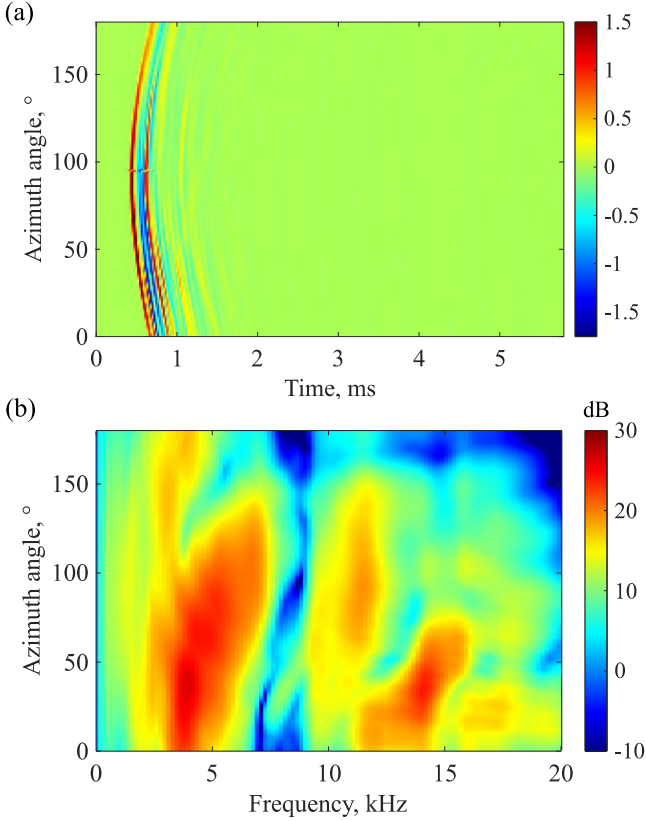

Fig. 3. Interpolated HRIRs and HRTFs of the left ear from the FABIAN dummy head and torso database at 0° elevation (horizontal plane) and 0° HATO: (a) HRIRs; (b) HRTFs.

was used, which provides highly detailed directional HRIRs for the FABIAN dummy head and torso across 11 head-above-torso orientations (HATO). The database includes measured HRIRs for azimuth angles ranging from 0° to 360° and elevation angles from –90° to 90°, sampled at 2° increments. Each HRIR consists of 256 samples ($K = 256$), and the measurements were conducted in an anechoic chamber at a 44.1 kHz sampling rate, covering the full audible frequency range from 20 Hz to 20 kHz. In this study, we used the measured HRIRs at 0° HATO as our target subject.

To simulate the continuous rotation of a speaker array using the measured HRIRs, we interpolated the azimuth resolution from 2° to 0.001°, reflecting the rotational speed of the speaker array set at 45°/s. For interpolation, we employed the inverse distance weighting method [48], which estimates HRIRs by weighting the nearest measured HRIRs inversely proportional to their angular distance from the target direction. We selected the interpolated HRIRs for the 12 speakers ($S = 12$) evenly spaced at 30° intervals along the elevation plane, ranging from -90° to 240°. The selected interpolated HRIRs were treated as ground-truth HRIRs to evaluate the accuracy of the proposed DNN-based HRIR identification by comparing the identified and true HRIRs. A 180° rotation scenario was simulated, resulting in $N = 176,400$ to cover the full 360° azimuth angle range with the 12 speakers. An example of the true HRIRs and head-related transfer functions (HRTFs) for the speaker at 0° elevation are illustrated in Fig. 3(a) and (b), respectively.

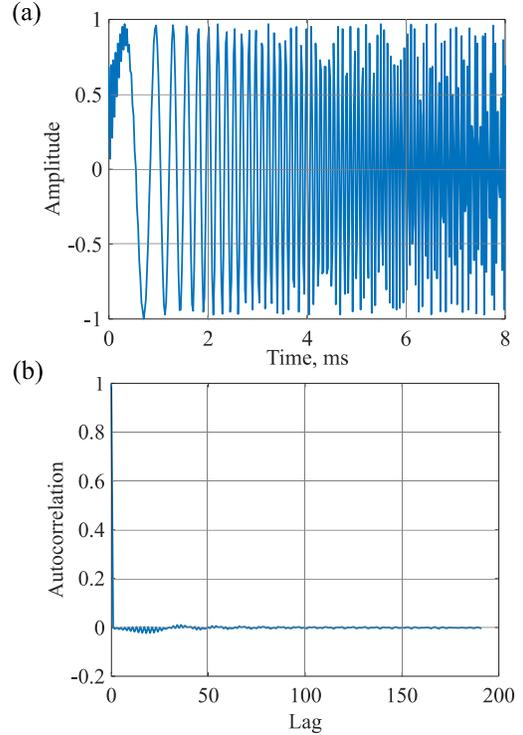

Fig. 4. Waveform and autocorrelation of perfect sweep used for the excitation signal of the continuously rotating speaker array in the simulation and experiment: (a) Wave form; (b) Autocorrelation.

The estimated HRIR length was selected to $\widetilde{K} = 192$, which was sufficient to capture all essential acoustic effects produced by the head and torso.

A perfect sweep signal was used as the excitation signal for the speaker array, as illustrated in Fig. 4. The perfect sweep, a sequence with an autocorrelation function of zero except at zero lag, has been shown in [35] to accelerate convergence speed and enhance robustness in measurement systems. A period length of the perfect sweep was set to $S\widetilde{K}$, and the excitation signal for the $s$-th speaker was circularly shifted by $(s - 1)\widetilde{K}$ to prevent cross-correlation between the excitation signals from different speakers. Using these excitation signals, binaural output signals for the left and right ears were generated according to (1). Note that the $K$ most recent samples from each speaker's excitation signal were used in the output signal generation process. at each time $n$. Additionally, white Gaussian noise with a variance of 0.01 was added to the output signal to simulate microphone noise, with the signal-to-noise ratio (SNR) set to 30 dB, representing a typical measurement environment.

Using this simulated data for the continuous rotation of the speaker array, we conducted the DNN-based HRIRs identification. While the identification can be performed for the full-sequence ($N = 176,400$), it was computationally intensive—requiring 36 hours. To mitigate this, the full-sequence was divided into 10 segments ($N = 17,640$ each), and identification was executed in parallel using 4 GPUs (NVIDIA RTX A6000). The DNN model, comprising 53 million parameters, completed the full-sequence identification in 9 hours.

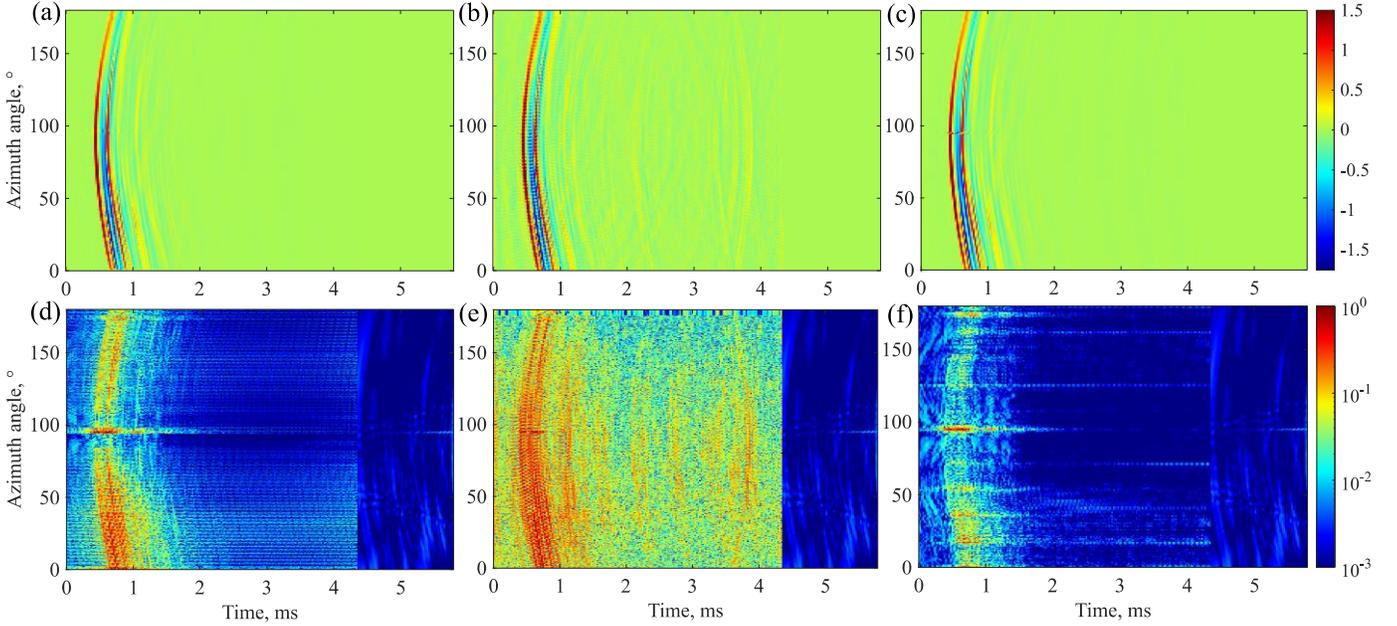

Fig. 5. Comparison of identified HRIRs on the horizontal plane using simulation data: (a)-(c) Identified HRIRs; (d)-(f) Absolute difference between true HRIRs and identified HRIRs; (a), (d) NLMS; (b), (e) Meta-AF; (c), (f) DNN-based HRIRs identification.

TABLE I
OBJECTIVE PERFORMANCE COMPARISON OF THE PROPOSED DNN-BASED HRIRS IDENTIFICATION, ANALYTICAL MODELS, META-AF, AND ABLATION CONDITIONS ON SIMULATION DATA IN TERMS OF NM AND LSD

| Methods | NM, dB | LSD, dB |
| --- | --- | --- |
| NLMS [24] | -18.45 | 3.56 |
| Kalman filter [29] | -16.89 | 4.83 |
| JO-NLMS [45] | -18.38 | 3.58 |
| Meta-AF [44] | -4.08 | 30.74 |
| w/o reset and update gate | -20.56 | 3.08 |
| Fixed normalization | -14.13 | 7.52 |
| $N/10$ sequence updating and optimization | 75.29 | 550.71 |
| $N/2$ sequence updating and optimization | -22.85 | 3.18 |
| **Proposed method** | **-25.58** | **1.74** |

*B. Results*

To objectively evaluate the performance of the DNN-based HRIRs identification, we employed normalized misalignment (NM) [49] to assess the accuracy of HRIRs identification in the time domain. NM is defined as:

$$\mathrm{NM} = \frac{1}{NS}\sum_{n=0}^{N}\sum_{s=1}^{S} 10\log_{10}\left(\frac{\|\mathbf{h}_{n,s}-\tilde{\mathbf{h}}_{n,s}\|^2}{\|\mathbf{h}_{n,s}\|^2}\right). \quad (9)$$

Where, $\mathbf{h}_{n,s}$ represents the interpolated HRIRs vector (i.e., the true HRIRs vector), and $\tilde{\mathbf{h}}_{n,s}$ denotes the identified HRIR vector from the identification process. Additionally, we used log spectral distortion (LSD) to validate the HRIRs identification in the frequency domain. LSD is expressed as:

$$\mathrm{LSD} = \sqrt{\frac{1}{NSI}\sum_{n=0}^{N}\sum_{s=1}^{S}\sum_{i=1}^{I}\left\{20\log_{10}\frac{|\mathbf{H}_{n,s}(f_i)|}{|\tilde{\mathbf{H}}_{n,s}(f_i)|}\right\}^2}, \quad (10)$$

where $\mathbf{H}_{n,s}(f_i)$ is the HRTFs vector corresponding to $\mathbf{h}_{n,s}$, and $i$ is the frequency bin index. The averages of NM and LSD for both left and right ears were used to evaluate overall performance.

Before computing these metrics, zero-padding was applied to the HRIRs identification results to match the length of $\tilde{\mathbf{h}}_{n,s}$ ($\tilde{K} = 196$) to that of $\mathbf{h}_{n,s}$ ($K = 256$). NM and LSD values were calculated for the DNN-based HRIRs identification and compared with those of conventional analytical models, including NLMS [24], Kalman filter [29], and JO-NLMS [45]. JO-NLMS introduces a variable step size and normalization factor to adaptively adjust these parameters based on $n$. We also compared with Meta-AF, DNN-based system identification with a training dataset. The Meta-AF model was trained in advance using measured HRIRs from 96 subjects in the HUTUBS database [37]. For construction of the training dataset for Meta-AF, HRIRs from HUTUBS database were interpolated to simulate the same 180° rotation scenario with a 45°/s speed.

Hyperparameters of the analytical models and Meta-AF were set as follows:
- NLMS: $\mu = 0.5$
- Kalman filter: process noise covariance matrix $= 10^{-7} \cdot \mathbf{I}$, measurement noise variance $= 0.01$
- JO-NLMS: same measurement noise variance as Kalman
- Meta-AF: adaptive filter unroll length $= 16$

The results are summarized in Table 1.

The proposed DNN-based HRIRs identification achieved a significant improvement in accuracy, maintaining NM of –

TABLE II
ABSOLUTE ITD ERROR BETWEEN IDENTIFIED ITDS AND TRUE ITDS AT 12 AZIMUTH ANGLES ON THE HORIZONTAL PLANE
(SCALED IN MICROSECONDS)

| Methods | 0° | 30° | 60° | 90° | 120° | 150° | 180° | 210° | 240° | 270° | 300° | 330° |
|---|---|---|---|---|---|---|---|---|---|---|---|---|
| NLMS [24] | 22.68 | 15.42 | 18.14 | 0.91 | 10.20 | 24.26 | 28.34 | 31.07 | 24.26 | 3.63 | 3.17 | 15.42 |
| Kalman filter [29] | 45.26 | 13.27 | 28.21 | 5.18 | 7.98 | 33.10 | 19.52 | 32.55 | 46.33 | 14.03 | 25.57 | 39.44 |
| JO-NLMS [45] | 23.71 | 18.10 | 17.33 | 11.43 | 32.57 | 20.20 | 27.07 | 17.81 | 29.66 | 20.03 | 15.32 | 19.70 |
| Meta-AF [44] | 77.78 | 156.01 | 4.08 | 44.67 | 20.41 | 435.61 | 9.98 | 109.75 | 80.73 | 223.13 | 288.21 | 36.28 |
| **Proposed method** | **8.34** | **1.59** | **8.39** | **3.17** | **2.49** | **5.67** | **5.13** | **3.63** | **0.91** | **3.17** | **6.35** | **1.13** |

25.58 dB and LSD of 1.74 dB. This performance is attributed to the complex neural network architecture with sequence-to-sequence learning, which effectively models the dynamic variations in HRIRs caused by the fast speaker array rotation. The DNN model successfully captures complex acoustic phenomena, including the head shadow effect and head and torso reflections. In contrast, JO-NLMS showed a slight degradation in performance, with NM worsening by 0.07 dB and LSD by 0.02 dB compared to NLMS. These results highlights the advantage of the DNN-based method, particularly its learnable normalization mechanism, which helps stabilize $\triangledown \text{ISE}(n)$ under rapid rotational conditions.

Fig. 5 compares the identified HRIRs for the left ear from the proposed method with those obtained using NLMS and Meta-AF. These two algorithms were selected as the delegate of the analytical models and DNN-based model, respectively. The proposed method significantly reduced identification errors in the critical time range of 0.1-1 ms, when key acoustic effects, such as head and pinna reflections, are most prominent. Additionally, while NLMS exhibited degraded accuracy in the 1.5-4 ms range due to reflections from measurement equipment (e.g., the speaker array), the proposed method maintained negligible errors. Although Meta-AF produced inaccurate HRIRs across nearly the entire time range, the proposed DNN method yielded consistently accurate results, benefiting from a network specifically tailored for HRIR identification. Note that any observed errors beyond 4.3 ms are attributed to zero-padding.

An ablation study was also conducted to assess the contributions of key components in the DNN model, as well as the whole-sequence update and optimization scheme. The results, summarized in Table 1, demonstrate that removing the GRU's reset and update gates while retaining the 3-layer FC network led to performance degradation, with NM and LSD worsening by 5 dB and 1.3 dB, respectively. This indicates that the FC network alone is insufficient to capture long-term temporal transitions in HRIRs due to the vanishing gradient problem. Furthermore, replacing the learnable normalization with a fixed normalization, i.e., dividing $\nabla \text{ISE}(n)$ by the power of $\mathbf{x}_{n,ele}$, as in NLMS, resulted in a sharp decline in accuracy (NM degraded by 11 dB and LSD by 6 dB). This emphasizes the critical role of learnable normalization in stabilizing the scale of $\nabla \text{ISE}(n)$ and controlling the update rate of $\tilde{\mathbf{h}}_{n,s}$ during the fast speaker rotation. Lastly, while the $N/2$ sequence updating and optimization scheme (i.e., updating $\tilde{\mathbf{h}}_{n,s}$ for $N/2$ sequence, followed by optimization of the DNN model) performed comparably to the whole sequence updating, the $N/10$ updating scheme led to overfitting and severe gradient explosion. Overall, the proposed method—combining the GRU structures, learnable normalization, and whole-sequence updating and optimization scheme—achieved the most accurate NM and LSD values, outperforming all other methods.

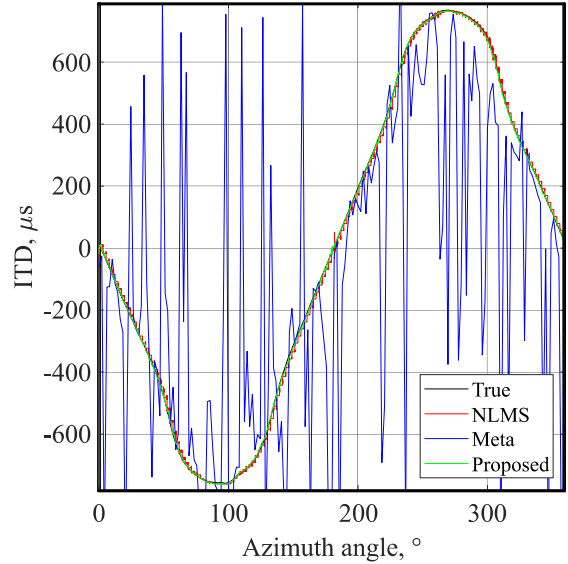

Fig. 6. Comparison of true ITDs and identified ITDs on the horizontal plane using simulation data. Results are shown for DNN-based HRIRs identification, NLMS, and Meta-AF.

To further verify that the identified HRIRs of the proposed method preserve accurate spatial localization cues, we performed analyses in both the time and frequency domains. In the time domain, interaural time difference (ITD) is a fundamental cue for azimuth localization. ITD is defined as [Xie]:

$$\text{ITD} = \arg\max_{\tau} \frac{\sum_{k=0}^{K-1} h_L(k) h_R(k-\tau)}{\sqrt{\sum_{k=0}^{K-1} h_L^2(k) \sum_{k=0}^{K-1} h_R^2(k)}}, \quad (11)$$

where $h_L$ and $h_R$ denote the HRIRs for the left and right ears, respectively. We compared the ITDs derived from the identified HRIRs to those of the true HRIRs across the horizontal plane, where ITD varies most significantly with azimuth angle. The results, shown in Fig. 6, reveal that the proposed method's identified ITDs closely matched the true ITDs across all azimuth angles, outperforming both NLMS and Meta-AF. For a quantitative comparison, we computed the absolute ITD error between the identified and true ITDs at 12 azimuth angles (0° to 330°, in 30° increments) in the horizontal plane. Table 2 summarizes these results. Considering that the just-noticeable difference (JND) of ITD is approximately 19 μs for frontal directions and 72 μs for lateral directions, the proposed method's errors remained consistently below 12 μs within perceptual thresholds. This confirms that the proposed model preserves accurate cues for azimuth localization.

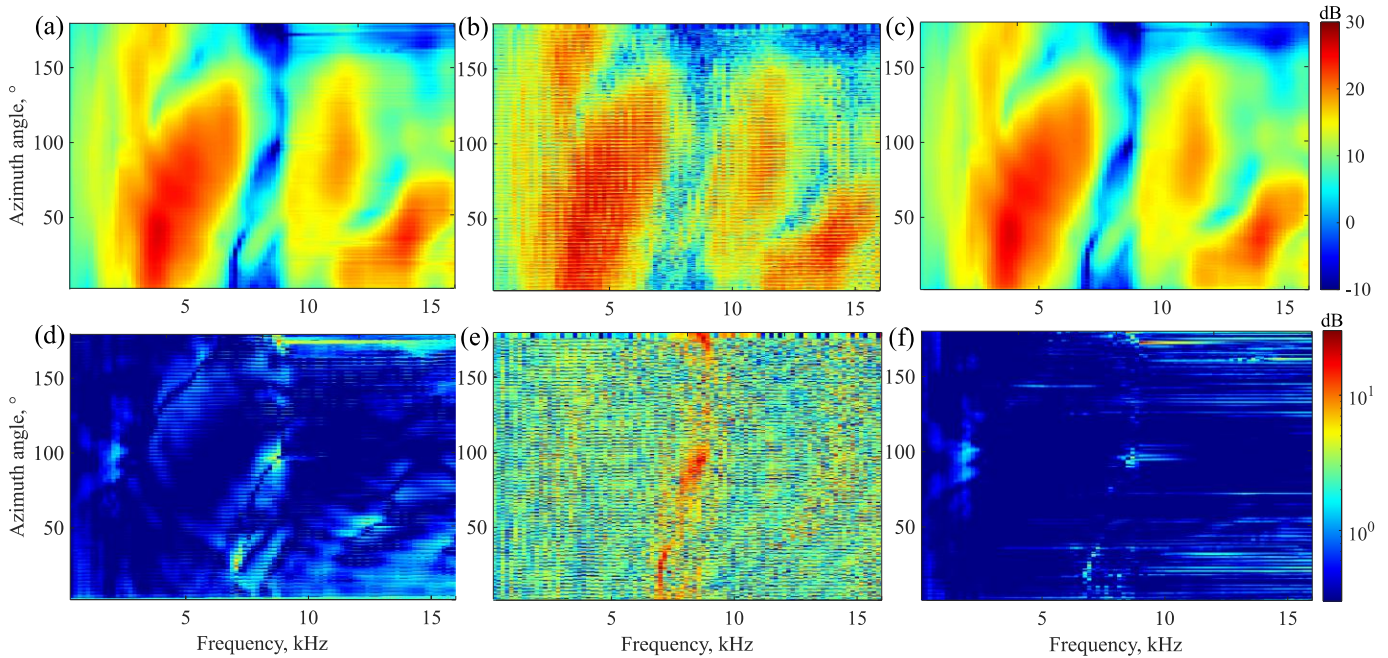

Fig. 7. Comparison of identified HRTFs on the horizontal plane using simulation data: (a)-(c) Identified HRTFs; (d)-(f) Absolute level difference between true HRTFs and identified HRTFs; (a), (d) NLMS; (b), (e) Meta-AF; (c), (f) DNN-based HRIRs identification.

In the frequency domain, monaural localization cues, particularly spectral patterns between 4-16 kHz, are mainly used for elevation localization [50]. The identified HRTFs for the left ear were compared with those from NLMS and Meta-AF. As shown in Fig. 7, the proposed method significantly outperformed the others in the 4-10 kHz range and achieved LSD values below 0.1 dB in frequencies below 7 kHz. Some distortion was observed above 13 kHz due to the fast rotation of speaker array causing excessive fluctuations in the high-frequency components. Nonetheless, the proposed method demonstrated high accuracy across the 4-16 kHz range overall, indicating reliable preservation of elevation localization cues.

To further demonstrate the preservation of monaural localization cues, Fig. 8 compares the identified and true HRTFs at several directions. For frontal directions ($\theta = 5°, \phi = 0°$) and ($\theta = 170°, \phi = 150°$), the proposed method accurately reconstructed spectral peaks and notches, while other methods showed notable distortion. For the upper left side direction ($\theta = 75°, \phi = 60°$), it successfully captured the deep spectral notch at 15 kHz, which results from interference between the direct wave and reflections off the concha wall. In the lower rear direction ($\theta = 15°, \phi = 240°$), some distortion was observed above 15 kHz, likely caused by frequent spectral fluctuations due to torso reflections and pinna diffraction associated with sounds arriving

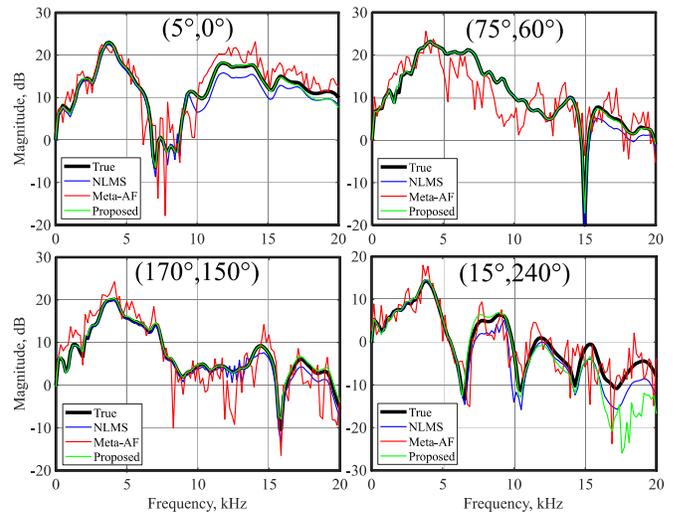

Fig. 8. Comparison of true HRTFs and identified HRTFs at several directions using simulation data. Results are shown for DNN-based HRIRs identification, NLMS, and Meta-AF. The title ($\theta,\phi$) denotes azimuth angle of $\theta$ and elevation angle of $\phi$ for sound source direction.

from lower rear side.

## VI. EXPERIMENT

### A. Experimental Set-up

We validated the robustness of the proposed DNN-based HRIR identification under experimental conditions using a custom-built speaker array and acoustic measurement setup. The experiment was conducted in an anechoic chamber with dimensions of 3.6 m × 3.6 m × 2.4 m and a cut-off frequency of 170 Hz. The frequency range of interest was set from 200

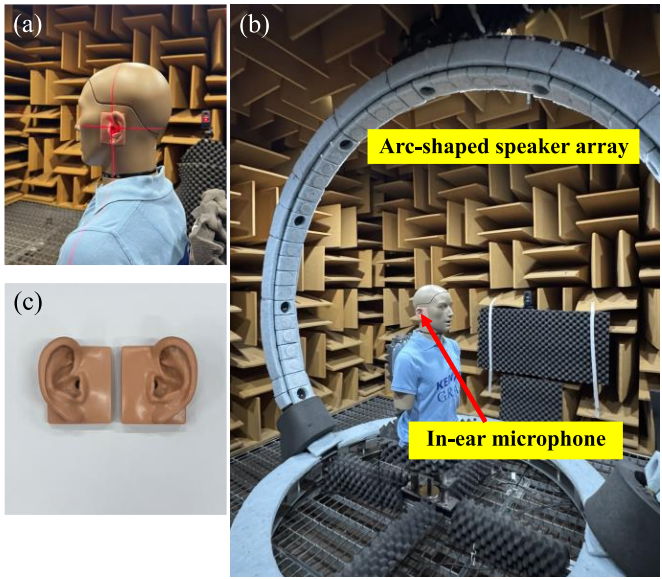

Fig. 9. Experimental setup for DNN-based HRIRs identification using the rotating arc-shaped speaker array: (a) Dummy head, GRAS 45BC KEMAR HATS; (b) Arc-shaped speaker array in an anechoic chamber; (c) Ear simulators, GRAS KB0065 and KB0066.

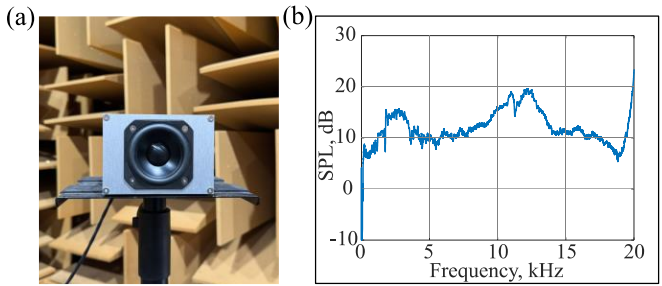

Fig. 10. Designed speaker module for the arc-shaped speaker array, along with its frequency response: (a) Speaker module; (b) Frequency response.

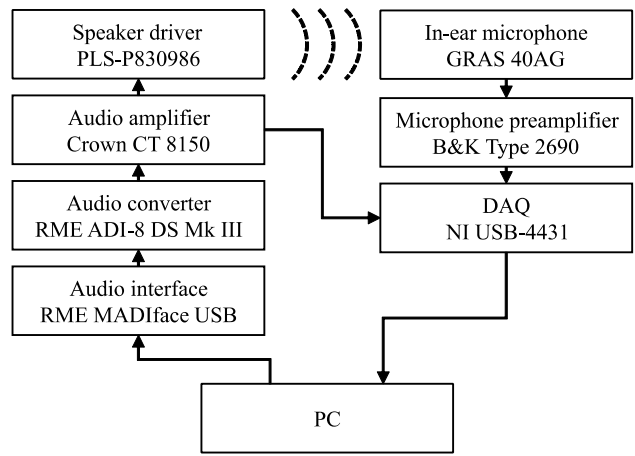

Fig. 11. Speaker and microphone system used in the experiment for DNN-based HRIRs identification.

Hz to 17 kHz, based on the effective frequency range for elevation localization (4-16 kHz) [50], the ITD-relevant frequency range (below 1.5 kHz) [51], and the anechoic chamber's cut-off frequency. A GRAS 45BC KEMAR HATS, equipped with Type KB0065 and KB0066 ear simulators, was used as the target subject to ensure experimental reproducibility and accurate evaluation of identification results.

The speaker system consisted of an arc-shaped array comprising 9 speakers ($S = 9$) positioned along the elevation plane from –30° to 210° in 30° increments, as shown in Fig. 9. The array had a radius of 1.1 m, approximating far-field HRTF conditions [52], and was mounted on a turntable to enable continuous rotation. Each speaker module incorporated a PLS-P830986 full-range speaker driver within an enclosure, with a −6 dB roll-off frequency of 119 Hz (Fig. 10), covering the frequency range of interest.

The estimated HRIR length was set to $\widetilde{K} = 340$, sufficient to capture all relevant acoustic effects from the head, torso, speaker array, and turntable.

To excite the speaker array, a perfect sweep signal with a period of $S\widetilde{K}$ was used. For each speaker $s$, the excitation signal was circularly shifted by $(s-1)\widetilde{K}$. These signals were delivered via an RME MADIface USB and RME ADI-648, routed to 2 D/A converters (RME ADI-8 DS), and amplified by 2 Crown CT 8150 amplifiers before being emitted through the speaker array. The emitted sounds were captured using GRAS 40AG in-ear microphones and amplified by a B&K Type 2690. All signals including the excitation signal from the first speaker (used to determine the onset time) and the binaural recordings from the microphones were simultaneously sampled at 34 kHz using an NI USB-4431. Fig. 11 provides an overview of the speaker and microphone setup.

The speaker array rotated at a constant speed of 36°/s, with minimal noise from the turntable motor. During a 5 s measurement, binaural signals were continuously captured as the array emitted sweep signals while rotating. The SNR of the experimental setup was 48 dB. A high-pass filter with a 200 Hz cut-off was applied to the recordings to match the defined frequency range of interest.

For DNN-based HRIRs identification, the full-sequence ($N = 170{,}000$) was divided into 10 segments ($N = 17{,}000$ each), and HRIRs identification was performed on each segment, as described in Section V.

To compensate for the measurement system's frequency responses (e.g., speaker and microphone frequency responses), origin transfer functions (OTFs) were measured using an in-ear microphone detached from the HATS. The identified HRTFs were then corrected by dividing them by the OTF. To further reduce the influence of acoustic reflections from surrounding equipment, a rectangular window limited HRIRs to the 0–6 ms range, accounting for early reflections caused by the speaker array and turntable.

The proposed DNN-based HRIRs identification method, incorporating the GRU structures, learnable normalization based on $\mathbf{x}_{n,ele}$, and whole-sequence updating and optimization scheme, was applied to the experimental data. For performance evaluation, ground-truth HRIRs of the KEMAR HATS were obtained using MESM, with the speaker array rotating stepwise from 0° to 359° in 1° azimuth increments. To compute NM and LSD, the true HRTFs, originally sampled at 51.2 kHz, were downsampled to 34 kHz to match the identified HRIRs. The length of true HRIRs was $K = 340$. Zero-padding was used to align HRIR lengths. Since the azimuth resolution of identified HRIRs differed from the true HRIRs, the identified HRIRs were interpolated at 1° azimuth intervals from 0° to 359° by selecting the closest available azimuth.

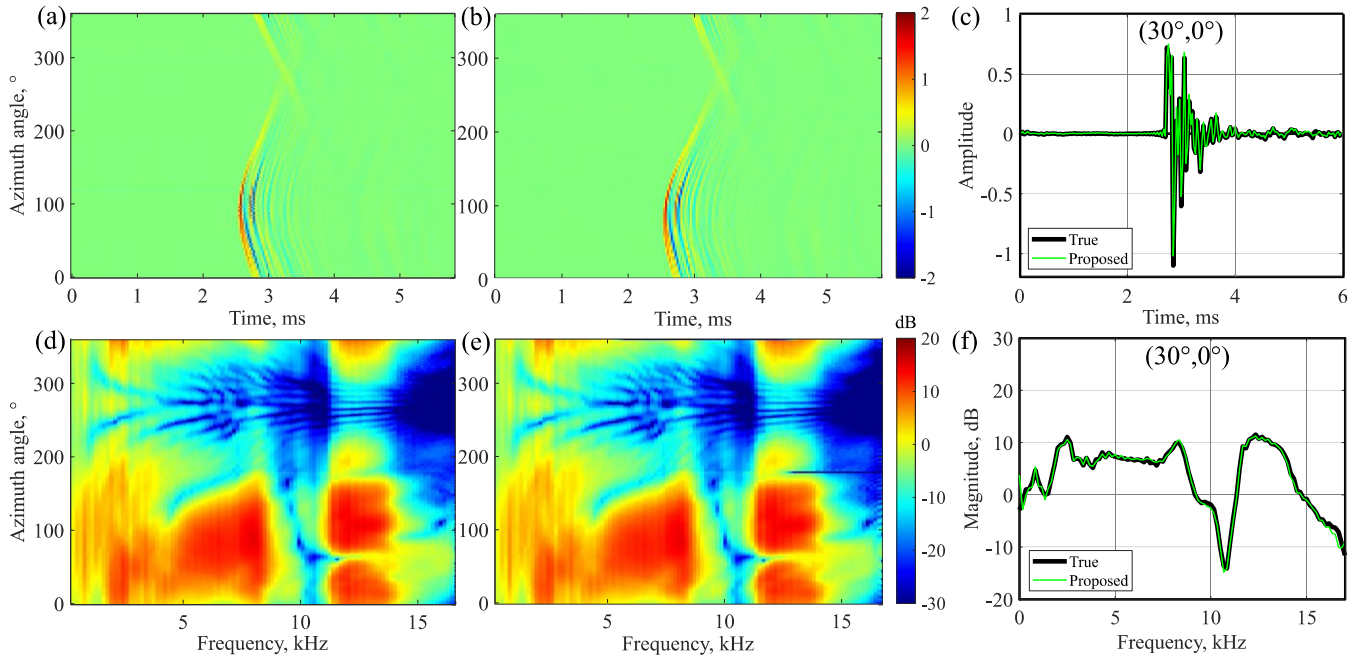

Fig. 12. Comparison of the true and identified HRIRs/HRTFs on the horizontal plane using experimental data: (a), (d) True HRIRs and HRTFs; (b), (e) Identified HRIRs and HRTFs from DNN-based HRIRs identification; (a)-(b) HRIRs on the horizontal plane; (d)-(e) HRTFs on the horizontal plane; (c) HRIR at azimuth 30°, elevation 0°; (f) HRIR at azimuth 30°, elevation 0°.

TABLE III
OBJECTIVE PERFORMANCE COMPARISONS OF THE PROPOSED DNN-BASED HRIRS IDENTIFICATION, ANALYTICAL MODELS, AND META-AF ON EXPERIMENTAL DATA IN TERMS OF NM AND LSD

| Methods | NM, dB | LSD, dB |
| --- | --- | --- |
| NLMS [24] | -16.81 | 4.70 |
| Kalman filter [29] | -12.34 | 8.27 |
| JO-NLMS [45] | -17.00 | 4.56 |
| Meta-AF [44] | -2.32 | 41.01 |
| **Proposed method** | **-23.21** | **1.98** |

As in Section V, the accuracy of the DNN-based HRIRs identification was compared with that of analytical models (NLMS, Kalman filter, and JO-NLMS) and Meta-AF. The trained Meta-AF's DNN model using the HUTUBS database in Section V was applied. Hyperparameters remained consistent with Section V, except for the measurement noise variance in the Kalman filter and JO-NLMS, which was adjusted based on measured noise in the experimental setup.

*B. Results*

Identification accuracy was evaluated using NM and LSD metrics, averaged across both ears. Table 3 summarizes the results, showing that the proposed DNN-based method achieved the best performance, with NM of –23.21 dB and LSD of 1.98 dB, outperforming the compared algorithms. These results confirm the method's robustness under experimental conditions. Its flexibility, owing to its neural network architecture, enables effective adaptation to varied conditions compared to analytical models. Meta-AF, in contrast, showed significantly reduced accuracy, an NM of –2.32 dB and an LSD of 41.01 dB, due to discrepancies between the experimental setup and the measurement conditions of HUTUBS database. Despite not requiring training data, the proposed method's sequence-to-sequence learning and full-sequence optimization enabled effective tracking of HRIR transitions. Nevertheless, minor performance degradation was observed relative to simulation due to experimental factors such as microphone noise and residual reflections from the setup.

Fig. 12 shows a comparison of identified and true HRIRs and HRTFs in both time and frequency domains, across the horizontal plane. Results from 0° and 180° elevation speakers were concatenated to present data across the full 360° azimuth range. The proposed method accurately reconstructed the critical 3-5 ms time range, corresponding to the direct path and early pinna/head reflections (given the 1.1 m speaker array radius). The reconstructed HRTFs exhibited an amplified magnitude in the ipsilateral direction (90° azimuth) and attenuation in the contralateral direction (270° azimuth), consistent with the true HRTFs and effectively demonstrating the proposed method's ability to capture the head shadow effect, a key cue for azimuthal localization [36]. Additional comparisons at single direction confirmed that the proposed method accurately estimated not only the amplitude and timing of the direct path and key reflections in HRIR, but also spectral cues in HRTF. Notably, it preserved the frequency and magnitude of prominent peaks and notches, which are essential for accurate elevation localization [53]. Overall, the proposed DNN-based HRIRs identification demonstrated precise preservation of spatial cue across azimuth and elevation, enabling accurate 3D sound localization when the identified HRIRs and HRTFs are applied in spatial audio synthesis.

## VII. CONCLUSION

This paper presented a novel DNN-based method for HRIRs identification, enabling accurate estimation of HRIRs using a continuously rotating speaker array within a short measurement time. To effectively track HRIR transitions during rapid speaker rotation, the proposed DNN architecture integrates GRU structures and FC networks to capture temporal dynamics in the HRIRs vector. A learnable normalization based on the speaker excitation signal, adaptively adjusts the scale of ISE gradient and the update rate of the HRIRs vector over time, thereby enhancing robustness. Furthermore, a whole-sequence updating and optimization scheme is introduced to enable optimization of the DNN model without the need for a training dataset, preventing overfitting and ensuring accurate accuracy across the whole sequence.

The effectiveness of the proposed method was first validated using simulated data from the FABIAN database under fast rotation conditions (45°/s). Evaluation using NM and LSD metrics demonstrated superior performance over conventional analytical and DNN-based approaches, achieving NM of –25.58 dB and LSD of 1.74 dB. The model accurately captured HRIR characteristics in the critical 0.1–1 ms time range, including reflection and diffraction effects from the head and pinna. ITD analysis confirmed that estimation errors remained below 12 μs, well within the JND threshold, supporting reliable azimuth localization. Frequency-domain analysis further validated the proposed method's ability to preserve monaural spectral cues in the 4-16 kHz range, which are essential for elevation localization.

To assess real-world applicability, the proposed method was also evaluated under experimental conditions using a custom-built arc-shaped speaker array rotating at 36°/s. The DNN-based approach achieved consistent identification accuracy, with NM of –23.21 dB and LSD of 1.98 dB, outperforming conventional approaches. The model successfully preserved key HRTF features, including spectral peaks and notches critical for elevation perception and head-shadow effects relevant to azimuthal localization. These results confirm the proposed method's robustness and flexibility across both simulated and experimental settings.

Future work will include subjective listening tests using VR environments to evaluate perceptual localization accuracy using HRIRs identified by the proposed method. These tests will help establish perceptually acceptable error bounds and further refine the model's performance. Additionally, while the current study segmented the full sequence to manage computational load, future developments will focus on improving the whole-sequence updating and optimization scheme. This will allow full-sequence processing in a computationally efficient manner, reducing training time while maintaining identification accuracy.

## REFERENCES

[1] D. N. Zotkin, R. Duraiswami, and L. S. Davis, "Rendering localized spatial audio in a virtual auditory space," *IEEE Trans. Multimedia*, vol. 6, no. 4, pp. 553-564, Aug. 2004.
[2] A. Londero et al., "Auditory and visual 3D virtual reality therapy for chronic subjective tinnitus: Theoretical framework," *Virtual Reality*, vol. 14, no. 2, pp. 143–151, Jun. 2010.
[3] A. Uribe-Quevedo, C. Arevalo, J. Villegas, and W. Sun, "Exploring 2D, 3D and spatial audio user interfaces in VR for reminiscence therapy," in *Proc. IEEE 12th Global Conf. Consum. Electron. (GCCE)*, Osaka, Japan, Oct. 2023, pp. 703–707.
[4] L. Shim et al., "Feasibility of virtual reality-based auditory localization training with binaurally recorded auditory stimuli for patients with single-sided deafness," *Clin. Exp. Otorhinolaryngol.*, vol. 16, no. 3, pp. 217–224, Aug. 2023.
[5] S. N. Yao, "Headphone-based immersive audio for virtual reality headsets," *IEEE Trans. Consum. Electron.*, vol. 63, no. 3, pp. 300–308, Aug. 2017.
[6] X. Hu, A. Song, Z. Wei, and H. Zeng, "StereoPilot: A wearable target location system for blind and visually impaired using spatial audio rendering," *IEEE Trans. Neural Syst. Rehabil. Eng.*, vol. 30, pp. 1621–1630, 2022.
[7] A. H. Kamkar-Parsi and M. Bouchard, "Instantaneous binaural target PSD estimation for hearing aid noise reduction in complex acoustic environments," *IEEE Trans. Instrum. Meas.*, vol. 60, no. 4, pp. 1141–1154.
[8] G.-T. Lee, S.-M. Choi, B.-Y. Ko, and Y.-H. Park, ''HRTF measurement for accurate sound localization cues,'' 2022, *arXiv:2203.03166*.
[9] F. Keyrouz, "Advanced binaural sound localization in 3-D for humanoid robots," *IEEE Trans. Instrum. Meas.*, vol. 63, no. 9, pp. 2098–2107, 2014.
[10] E. M. Wenzel, M. Arruda, D. J. Kistler, and F. L. Wightman, "Localization using nonindividualized head-related transfer functions," *J. Acoust. Soc. Am.*, vol. 94, no. 1, pp. 111–123, 1993.
[11] C. Jenny and C. Reuter, "Usability of individualized head-related transfer functions in virtual reality: Empirical study with perceptual attributes in sagittal plane sound localization," *JMIR Serious Games*, vol. 8, no. 3, Jul. 2020.
[12] T. Huttunen et al., "Simulation of the transfer function for a head-and-torso model over the entire audible frequency range," presented at the *Proc. 2007 Int. Congr. Acoustics (ICA)*, Madrid, Spain, Sep. 2007.
[13] Y. Kahana and P. A. Nelson, "Numerical modelling of the spatial acoustic response of the human pinna," *J. Sound Vib.*, vol. 292, no. 1–2, pp. 148–178, Apr. 2006.
[14] B.-Y. Ko, G.-T. Lee, H. Nam, and Y.-H. Park, "PRTFNet: HRTF individualization for accurate spectral cues using a compact PRTF," *IEEE Access*, vol. 11, pp. 96119–96130, 2023.
[15] T. Chen, T. Kuo, and T. Chi, "Autoencoding HRTFs for DNN based HRTF personalization using anthropometric features," in *Proc. IEEE Int. Conf. Acoustics, Speech, Signal Process. (ICASSP)*, Singapore, May 2019.
[16] J. G. Richter and J. Fels, "On the influence of continuous subject rotation during high-resolution head-related transfer function measurements," *IEEE/ACM Trans. Audio Speech Lang. Process.*, vol. 27, no. 4, pp. 730–741, Apr. 2019.
[17] P. Majdak, P. Balazs, and B. Laback, "Multiple exponential sweep method for fast measurement of head-related transfer functions," *J. Audio Eng. Soc.*, vol. 55, no. 11, pp. 623–637, Nov. 2007.
[18] K. Fukudome, T. Suetsugu, T. Ueshin, R. Idegami, and K. Takeya, "The

fast measurement of head-related impulse responses for all azimuthal directions using the continuous measurement method with a servo-swiveled chair," *Appl. Acoust.*, vol. 68, no. 8, pp. 864–884, Aug. 2007.
[19] J.-G. Richter, "Fast measurement of individual head-related transfer functions," Ph.D dissertation, RWTH Aachen, Aachen, Germany, 2019.
[20] V. Pulkki, "HRTF measurements with a continuously moving loudspeaker and swept sines," in Proc. 128th Audio Eng. Soc. Conv., 2010.
[21] G. Enzner, "3D-continuous-azimuth acquisition of head-related impulse responses using multi-channel adaptive filtering," in *Proc. IEEE Workshop Appl. Signal Process. Audio Acoust.*, 2009.
[22] M. Fallahi, F. Brinkmann, and S. Weinzierl, "Simulation and analysis of measurement techniques for the fast acquisition of head-related transfer functions," in *Proc. German Annu. Conf. Acoust. (DAGA)*, Nuremberg, Germany, Sep. 7–12, 2015.
[23] S. Li and J. Peissig, "Measurement of head-related transfer functions: A review," *Appl. Sci.*, vol. 10, no. 14, p. 5014, Jul. 2020.
[24] E. L. Tan, S. Peksi, and W. S. Gan, "Implementing continuous HRTF measurement in near-field," in *Proc. IEEE Int. Conf. Acoust., Speech Signal Process. (ICASSP)*, 2023.
[25] G. Enzner, "Analysis and optimal control of LMS-type adaptive filtering for continuous-azimuth acquisition of head-related impulse responses," in *Proc. IEEE Int. Conf. Acoust., Speech Signal Process. (ICASSP)*, Mar. 2008, pp. 393–396.
[26] S. Li, A. Tobbala, and J. Peissig, "Towards mobile 3D HRTF measurement," in *Proc. 148th Audio Eng. Soc. Conv.*, May 2020.
[27] M. Fallahi, "Simulation and analysis of measurement techniques for the fast acquisition of individual head-related transfer functions," M.S. thesis, Tech. Univ. Berlin, Berlin, Germany, 2014.
[28] T. Kabzinski and P. Jax, "A flexible framework for expectation maximization-based MIMO system identification for time-variant linear acoustic systems," *IEEE Open J. Signal Process.*, vol. 5, pp. 112–121, 2023.
[29] T. Kabzinski and P. Jax, "Towards faster continuous multi-channel HRTF measurements based on learning system models," in *Proc. IEEE ICASSP*, 2022, pp. 436–440.
[30] F. Brinkmann, A. Lindau, S. Weinzierl, G. Geissler, and S. van de Par, "A high-resolution head-related transfer function database including different orientations of head above the torso," Tech. Univ. Berlin, Berlin, Germany, Tech. Rep., 2019.
[31] G. Bertolini and D. Straumann, "Moving in a moving world: A review on vestibular motion sickness," *Front. Neurol.*, Feb. 2016.
[32] F. Carrick, G. Pagnacco, E. Oggero, and D. Barton, "The effects of whole body rotations in the pitch and yaw planes on postural stability," *Funct. Neurol.*, 2011.
[33] L. R. Rabiner and R. W. Schafer, "Introduction to digital speech processing," *Found. Trends Signal Process.*, vol. 1, no. 1–2, pp. 1–194, 2007.
[34] C. Antweiler and P. Vary, "Sequential and direct access of head-related transfer functions (HRTFs) for quasi-continuous angular positions," in *Proc. 19th Eur. Signal Process. Conf. (EUSIPCO)*, Aug. 2011, pp. 1648–1652.
[35] C. Antweiler, A. Telle, P. Vary, and G. Enzner, "Perfect-sweep NLMS for time-variant acoustic system identification," in *Proc. IEEE Int. Conf. Acoust., Speech Signal Process. (ICASSP)*, Mar. 2012, pp. 517–520.
[36] B. Xie, *Head-Related Transfer Function and Virtual Auditory Display*. Plantation, FL, USA: J. Ross Publ., 2013.
[37] F. Brinkmann *et al.*, "A cross-evaluated database of measured and simulated HRTFs including 3D head meshes, anthropometric features, and headphone impulse responses," *J. Audio Eng. Soc.*, vol. 67, no. 9, pp. 705–718, 2019.
[38] W. Zhang, T. D. Abhayapala, R. A. Kennedy, and R. Duraiswami, "Insights into head-related transfer function: Spatial dimensionality and continuous representation," *J. Acoust. Soc. Am.*, vol. 127, no. 4, pp. 2347–2357, Apr. 2010.
[39] G. D. Romigh, D. S. Brungart, R. M. Stern, and B. D. Simpson, "Efficient real spherical harmonic representation of head-related transfer functions," *IEEE J. Sel. Topics Signal Process.*, vol. 9, no. 5, pp. 921–930, Aug. 2015.
[40] L. Russell, R. Goubran, and F. Kwamena, "Posture detection using sounds and temperature: LMS-based approach to enable sensory substitution," *IEEE Trans. Instrum. Meas.*, vol. 67, no. 7, pp. 1543–1554, Jul. 2018.
[41] Y. Wu *et al.*, "Adaptive filtering improved apnea detection performance using tracheal sounds in noisy environment: A simulation study," *Biomed. Res. Int.*, vol. 2020, 2020.
[42] B. Balasingam and K. R. Pattipati, "On the identification of electrical equivalent circuit models based on noisy measurements," *IEEE Trans. Instrum. Meas.*, vol. 70, 2021.
[43] C. K. Correa, S. Li, and J. Peissig, "Analysis and comparison of different adaptive filtering algorithms for fast continuous HRTF measurement," in *Proc. Fortschr. Akust. (DAGA)*, Kiel, Germany, Mar. 6–9, 2017.
[44] J. Casebeer, N. J. Bryan, and P. Smaragdis, "Meta-AF: Meta-learning for adaptive filters," *IEEE/ACM Trans. Audio Speech Lang. Process.*, vol. 31, pp. 355–370, 2023.
[45] S. Ciochină, C. Paleologu, and J. Benesty, "An optimized NLMS algorithm for system identification," *Signal Process.*, vol. 118, pp. 115–121, Jan. 2016.
[46] S. Ruder, "An overview of gradient descent optimization algorithms," 2016, *arXiv:1609.04747*.
[47] D. P. Kingma and J. Ba, "Adam: A method for stochastic optimization," 2014, *arXiv:1412.6980*.
[48] F. Brinkmann, R. Roden, A. Lindau, and S. Weinzierl, "Audibility and interpolation of head-above-torso orientation in binaural technology," *IEEE J. Sel. Topics Signal Process.*, vol. 9, no. 5, pp. 931–942, Aug. 2015.
[49] J. Benesty, H. Rey, L. R. Vega, and S. Tressens, "A nonparametric VSS NLMS algorithm," *IEEE Signal Process. Lett.*, vol. 13, no. 10, pp. 581–584, Oct. 2006.
[50] J. Hebrank and D. Wright, "Spectral cues used in the localization of sound sources on the median plane," *J. Acoust. Soc. Am.*, vol. 56, no. 6, pp. 1829–1834, 1974.
[51] L. Rayleigh, "XII, on our perception of sound direction," London, Edinburgh, Dublin Philos. Mag. J. Sci., vol. 13, no. 74, pp. 214–232, 1907.
[52] H. Gamper, "Head-related transfer function interpolation in azimuth, elevation, and distance," *J. Acoust. Soc. Am.*, vol. 134, no. 6, pp. EL547–EL553, Dec. 2013.
[53] K. Iida, M. Itoh, A. Itagaki, and M. Morimoto, "Median plane localization using a parametric model of the head-related transfer function based on spectral cues," *Appl. Acoust.*, vol. 68, no. 8, pp. 835–850, Aug. 2007.

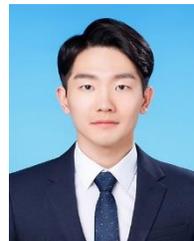

**BYEONG-YUN KO** received the B.S. degree in ship architecture and ocean engineering from Inha University, South Korea, in 2019, and the M.S. degree in mechanical engineering from KAIST, South Korea, in 2021. He is currently pursuing the Ph.D. degree in mechanical engineering at the same institute. His research interest includes the development of deep learning-based individualization and measurement of HRTF and spatial audio rendering technique considering the human auditory perception.

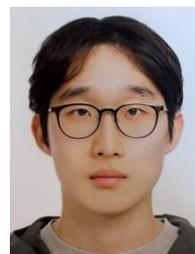

**DEOKKI MIN** received B.S. degree in mechanical engineering from Yonsei Univ. in 2022 and M.S. degree in mechanical engineering from Korea Advanced Institute of Science and Technology (KAIST), Daejeon, Korea, in 2024. He is currently working as an intern researcher at Korea Institute of Science and Technology (KIST). His interests include auditory system modeling, with a focus on acoustic signal processing, interpretable deep learning model, and neuroscientific theories related to auditory perception.

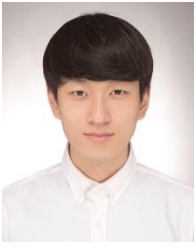
**HYEONUK NAM** (Member, IEEE) received the B.S. and M.S. degrees in mechanical engineering from Korea Advanced Institute of Science and Technology, Daejeon, Korea, in 2018 and 2020 respectively. He is currently pursuing the Ph.D. degree in mechanical engineering at the same institute. His research interest includes sound event detection, sound localization, automatic speech recognition, and speech dereverberation.

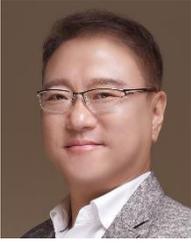
**YONG-HWA PARK** (Member, IEEE) received BS, MS, and PhD in Mechanical Engineering from KAIST in 1991, 1993, and 1999, respectively. In 2000, he joined to Aerospace Department in University of Colorado at Boulder as a research associate. From 2003-2016, he worked for Samsung Electronics in Visual Display Division and Samsung Advanced Institute of Technology (SAIT) as a Research Master in the field of micro-optical systems with applications to imaging and display systems. From 2016, he joined to KAIST as an associate professor of NOVIC+ (Noise & Vibration Control Plus) at department of Mechanical Engineering devoting to researches on vibration, acoustics, vision sensors and recognitions for human-machine interactions. His research fields include structural vibration; event/condition recognition from sound and vibration signatures utilizing AI; blood pressure and health monitoring sensors; and 3D sensors, lidar for motion measurements. He works as a conference chair of MOEMS and miniaturized systems in SPIE Photonics West since 2013. He is a board member of KSME, KSNVE, KSPE, and SPIE.